\documentclass[copyright,creativecommons]{eptcs}
\providecommand{\event}{QAPL 2012} 
\usepackage{breakurl}              
\usepackage{amssymb}
\usepackage{amsmath}
\usepackage{graphicx}
\usepackage{rotating}
\usepackage{stmaryrd}
\usepackage{url}
\usepackage[mathscr]{eucal}

\newcommand{\MTC}{\mathit{MTC}}

\newcommand{\Input}{\mathit{Input}}
\newcommand{\Output}{\mathit{Output}}
\newcommand{\Buffer}{\mathit{Buffer}}
\newcommand{\Gen}{\mathit{Generate}}
\newcommand{\Drop}{\mathit{Drop}}
\newcommand{\KeepI}{\mathit{KeepI}}
\newcommand{\KeepG}{\mathit{KeepG}}
\newcommand{\Rem}{\mathit{Remove}}
\newcommand{\Node}{\mathit{Node}}
\newcommand{\Level}{\mathit{Level}}
\newcommand{\TotalIG}{\mathit{TotalIG}}
\newcommand{\TotalD}{\mathit{TotalD}}

\newcommand{\CInput}{\mathit{ConI}}
\newcommand{\COutput}{\mathit{ConO}}
\newcommand{\CGen}{\mathit{ConG}}
\newcommand{\CDrop}{\mathit{ConD}}
\newcommand{\CRem}{\mathit{ConR}}
\newcommand{\CLink}{\mathit{ConL}}
\newcommand{\CStream}{\mathit{ConStream}}
\newcommand{\CNode}{\mathit{ConNode}}
\newcommand{\CUni}{\mathit{ConUni}}

\newcommand{\CTidy}{\mathit{ConTidy}}
\newcommand{\off}{\mathit{off}}

\def\S{\mbox{\large $\rhd\!\!\!\!\!\lhd$}}
\def\rmdef{\stackrel{\mbox{\em {\tiny def}}}{=}}
\def\Chi{\mbox{\Large $\chi$}}

\newcommand{\sync}[1]{\raisebox{-1.0ex}{$\;\stackrel{\S}{\scriptscriptstyle
#1}\,$}}

\newcommand{\syncstar}{\smash{\sync{\rule{0pt}{2.8pt}}}\hspace*{-8.8pt}\raisebox{-3.3pt}[0pt][0pt]{$\scriptscriptstyle *$}\hspace*{8pt}}

\newtheorem{definition}{Definition}

\newcommand{\ev}[1]{\underline{\mathrm{#1}}}
\newcommand{\sev}[1]{\overline{\mathrm{#1}}}
\newcommand{\ssev}[1]{\mathrm{#1}}
\newcommand{\Ev}{\mathcal{E}}
\newcommand{\ec}{\mathop{\mathit{ec}}}
\newcommand{\EC}{\mathit{EC}}
\newcommand{\iname}[1]{\mathit{#1}}
\newcommand{\const}{\mathit{const}}
\newcommand{\Ac}{\mathcal{A}}
\newcommand{\iv}{\mathop{\mathit{iv}}}
\newcommand{\ID}{\mathit{ID}}
\newcommand{\IN}{\mathit{IN}}
\newcommand{\IT}{\mathit{IT}}

\newcommand{\assign}[1]{\llbracket #1 \rrbracket}
\newcommand{\W}{\mathcal{W}}
\newcommand{\V}{\mathcal{V}}

\newcommand{\ConSys}{\mathit{ConSys}}
\newcommand{\Con}{\mathit{Con}}
\newcommand{\CSys}{\mathcal{C}_\mathit{Sys}}

\newcommand{\bbR}{\mathbb{R}}
\newcommand{\pc}{{:}}

\title{Hybrid performance modelling of opportunistic networks}
\author{Luca Bortolussi$^1$ \qquad Vashti Galpin$^2$ \qquad Jane Hillston$^2$
\institute{$^1$ Department of Mathematics and Computer Science,
University of Trieste}
\institute{$^2$ Laboratory for Foundations of Computer Science, 
School of Informatics, University of Edinburgh}
\email{luca@dmi.units.it, Vashti.Galpin@ed.ac.uk, Jane.Hillston@ed.ac.uk}}

\def\titlerunning{Hybrid performance modelling of opportunistic networks}
\def\authorrunning{L. Bortolussi, V. Galpin \& J. Hillston}

\begin{document}

\maketitle

\begin{abstract}
We demonstrate the modelling of opportunistic networks using the
process algebra stochastic HYPE. Network traffic is modelled as
continuous flows, contact between nodes in the network is modelled
stochastically, and instantaneous decisions are modelled as discrete
events. Our model describes a network of stationary video sensors
with a mobile ferry which collects data from the sensors and delivers
it to the base station. We consider different mobility models and
different buffer sizes for the ferries.  This case study illustrates
the flexibility and expressive power of stochastic HYPE. We also
discuss the software that enables us to describe stochastic HYPE
models and simulate them.  
\end{abstract}

\section{Introduction}
\label{sec:intro}

Hybrid behaviour can arise in widely varying different contexts,
both engineered and natural, pure and abstracted.  Such systems
have elements which are subject to continuous change, interleaved
with discrete events which may change the elements themselves as
well as their mode of evolution. The continuous aspect of the
behaviour may be \emph{pure} in that it is a physical entity which
has continuous values, such as temperature or pressure, or may be
\emph{abstracted} as an approximation of a discrete quantity such
as concentrations of biochemical species within a cell.  Thus
examples of hybrid systems include thermostatically controlled
heating systems and genetic regulatory networks, such as the
repressilator \cite{ElowEL:00a,GHB08a}.

In this paper we consider an engineered system with abstract
continuity: an \emph{opportunistic} 
network \cite{HuanHLT:08a,PeluPPC:06a}.  In such a network, nodes
experience periods of disconnectedness, during which they nevertheless
may accumulate traffic in the form of packets.  Sporadic connectivity
is provided by occasional proximity of other nodes.  Such connectivity
is then exploited to further the progress of packets towards their
destination (hence the term \emph{opportunistic}).  There are many
interesting questions about performance and capacity planning for
such networks, but a detailed discrete state representation in which
all packets are treated individually can rapidly exceed feasible
analysis and can also be expensive in terms of time.  Instead, here,
we abstract the traffic to be a fluid quantity rather than discrete
packets and model the system as a stochastic hybrid system.

Process algebras have a long-established history of use for
compositional modelling and analysis of systems with concurrent
behaviour.   Moreover, when extended with stochastic variables to
represent duration and relative probability of events, they have
been successfully applied to performance modelling and other forms
of quantified analysis.  HYPE is one of several recently defined
process algebras which extend this capability to the modelling of
hybrid systems \cite{KhadK:06a,GalpGBH:09a}.  In HYPE the focus is
on a fine-grained compositionality, in which all the \emph{influences}
or \emph{flows} which impact the continuous variables in the system
are modelled explicitly.  Importantly, addressing the modelling in
this style removes the need to include ordinary differential equations
(ODEs) governing the continuous evolution of such variables in the
syntax of the model.  Instead the dynamic behaviour emerges, via
the semantics of the language, when the components are composed.
Moreover, this fine-grained approach gives the language more
expressiveness, than say, hybrid automata, as recently demonstrated
in \cite{HYPE-journal}.  Furthermore, the use of flows as the basic
elements of model construction has advantages such as ease and
simplification of modelling. This approach assists the modeller by
allowing them to identify smaller or local descriptions of the model
and then to combine these descriptions to obtain the larger system.

In the original definition of HYPE, discrete actions  were termed
\emph{events} and these were always considered to be \emph{instantaneous}
although they could be subject to an activation condition determining
just when the instantaneous jump  would occur.  A distinction was
made between \emph{urgent} and \emph{non-urgent} events.  Most
activation conditions are expressed in terms of conditions on the
evolving values of continuous variables and urgent events are
triggered immediately when such conditions become true.  In contrast,
non-urgent events were not tied to the continuous state of the
system (denoted by the undefined activation condition $\bot$) and
could occur randomly at some unspecified time in the future.  A
recent extension of HYPE \cite{BortBGH:11a} refined this notion of
non-urgent events, by introducing \emph{stochastic events}.  These
events have an activation condition that is a random variable,
capturing the probability distribution of the time until the event
occurs.  Thus these events still occur non-deterministically,
but they are now quantified and so the models admit quantitative
analysis.  In order to carried out this type of analysis, a novel
software tool has been developed which can simulate stochastic HYPE
models.

This extended HYPE is ideal for modelling opportunistic networks
in which we wish to study the emergent properties when nodes establish
contact intermittently, but according to some probability distribution.
In other words, we have some expectation of the frequency with which
connections are formed, rather than admitting the possibility that
this can be indefinitely postponed, as would be the case with
non-urgent events. This is more realistic since the intermittent
connectivity is usually provided by nodes embedded in vehicles which
make regular visits.

The case study presented in this paper consists of stationary nodes
that record multimedia data, specifically video, and a mobile node
on a vehicle that collects the data and delivers it to a stationary
base station. Since a characteristic of multimedia is very high
data volume, this scenario is particularly appropriate for the fluid
approach that we take here. We are specifically interested in two
parameters of the model: how much data storage is required for the
ferry and how often should it interact with the video sensors.

The rest of this paper is organised as follows. In Section~\ref{sec:sHYPE}
we briefly recall the basic notions of stochastic HYPE by means of
a running example. This includes an account of the novel software
tool which has been developed to simulate stochastic HYPE models.
Next, in Section~\ref{sec:Opportunistic} we give a more detailed description
of opportunistic networks, and the particular system we are
considering.  Section~\ref{sec:Model} presents a general framework
for describing opportunistic networks, presents the stochastic HYPE
model of the case study and the results of its analysis.  Finally,
Sections~\ref{sec:relatedWork} and \ref{sec:furtherResearch} discuss
related work, future research and draw final conclusions.

\section{Stochastic HYPE}
\label{sec:sHYPE}

In this section we present the definition  of stochastic HYPE
\cite{BortBGH:11a} and introduce a small example to illustrate the
definition. More details about the language can be found
in~\cite{BortBGH:11a,GalpGBH:09a,HYPE-journal}.  We consider a basic
model of a network node with a buffer, which can receive packets
from an input channel and send packets to an output channel. We
assume that the number of packets that travel through the node and
that are stored in the buffer is large, hence we describe them as
a fluid quantity.  Received packets are stored in the buffer, waiting
to be sent. We allow reception and sending of packets to happen
concurrently, but it is equally simple to enforce a mutually exclusive
send/receive policy.  We also assume that uplinks and downlinks are
not always working, but they are activated and deactivated depending
on the availability of a connection. These events are described as
stochastic, with firing times governed by exponential distributions.
Finally, incoming traffic has to be stopped if the buffer becomes
full and outgoing traffic has to be stopped when the buffer is
empty.

\begin{figure}[!t]

{\renewcommand{\arraystretch}{1.00}
$$\begin{array}{rcl}
\Buffer & \rmdef & Sys \sync{M}
\ev{init}.Con \quad \text{with} 
\quad M = \{ \ev{init}, \sev{on}_{in},
\sev{off}_{in}, \sev{on}_{out},
\sev{off}_{out}, \ev{empty}, \ev{full} \}.
\\
& & \\
Sys & \rmdef & \Input \sync{\{\ev{init}\}} \Output \\
& & \\
\Input & \rmdef & 
 \sev{on}_{in}\pc(in,r_{in},\const).\Input +
 \sev{off}_{in}\pc(in,0,\const).\Input +\\
 & & \ev{full}\pc(in,0,\const).\Input +
 \ev{init}\pc(in,0,\const).\Input               \\
\Output & \rmdef &   
 \sev{on}_{out}\pc(out,-r_{out},\const).\Output +
 \sev{off}_{out}\pc(out,0,\const).\Output +\\
 & &  \ev{empty}\pc(out,0,\const).\Output +
 \ev{init}\pc(out,0,\const).\Output \\
\\
Con & \rmdef & Con_{in} \sync{\emptyset} Con_{out}\\
Con_{in} & \rmdef & \sev{on}_{in}.Con_{in}' \quad \Con_{in}' \:\rmdef\:
\sev{off}_{in}.Con_{in} + \ev{full}.Con_{in}\\
Con_{out} & \rmdef & \sev{on}_{out}.Con_{out}' \quad \Con_{out}' \:\rmdef\:
\sev{off}_{out}.Con_{out} + \ev{empty}.Con_{out}\\
\end{array}$$

$$\begin{array}{rclrcl}
\iv(in) & = & B & \iv(out) & = & B \\
\\
\ec(\ev{init}) & = & \multicolumn{1}{l}{(true,B'=b_0)} \\
\ec(\sev{on}_{in}) & = & (k_{in}^{on},true) \quad & \ec(\sev{off}_{in})
& = & (k_{in}^{\off},true) \\
\ec(\sev{on}_{out}) & = & (k_{out}^{on},true) \quad &
\ec(\sev{off}_{out}) & = & (k_{out}^{\off},true) \\
\ec(\ev{full}) & = & (B = max_B,true) \quad & \ec(\ev{empty}) & = & (B = 0,true) \\
\end{array}$$
}
\caption{Simple network node model in stochastic HYPE.}\label{simpleNetwork}
\end{figure}

HYPE modelling is centred around the notion of \emph{flow}, which
is intended here as some sort of influence continuously modifying
one variable. Both the strength and form of a flow can be changed
by \emph{events}.  In our example, there are two flows modifying
the buffer level, modelled by the continuous variable $B$, namely
reception and sending of packets.  Flows are described by the
\emph{uncontrolled system}, a composition of several sequential
subcomponents, each modelling how a specific flow is changed by
events. For instance, in Figure~\ref{simpleNetwork}, the subcomponent
$\Input$ describes the inflow of packets in the buffer. This
subcomponent reacts to four events: $\sev{on}_{in}$ and $\sev{off}_{in}$,
modelling the activation and deactivation of the uplink; $\ev{full}$,
modelling the suspension of incoming traffic due to the buffer
becoming full; and $\ev{init}$, the first event that sets the initial
value of the influence. The tuple $(in,r_{in},\const)$ following
event $\sev{on}_{in}$, is called an \emph{activity} or an
\emph{influence} and describes how the input affects the buffer
level when it is in effect: $in$ is the name of the influence, which
provides a link to the target variable of the flow ($B$ in our
example), $r_{in}$ is the strength of the influence and $\const$
is the influence type, identifying the functional form of the flow
(which is specified separately by the interpretation $\assign{\const}
= 1$). When the input is switched off, the influence $(in,r_{in},\const)$
is replaced by $(in,0,\const)$ i.e. the influence strength of the
input becomes zero. The other subcomponent affecting buffer level
is the output component, modelling the sending of packets.  States
of a HYPE model are collections of influences, one for each influence
name, defining a set of ordinary differential equations describing
the continuous evolution of the system. For instance, $(in,r_{in},\const)$
contributes to the ODE of $B$ with the summand
$r_{in}\assign{\const}=r_{in}$.

The controller $Con$, instead, is used to impose causality on events,
reflecting natural constraints or design choices.  For instance,
$\Con_{in}$ models the fact that the reception of packets can be
turned off only after being turned on. Furthemore, it describes
termination of the input if the buffer becomes full, but only if
the uplink is active.

Events in stochastic HYPE are of two kinds, either \emph{stochastic}
or \emph{deterministic}.  Deterministic events $\ev{a}\in \Ev_d$
happen when certain conditions are met by the system.  These
\emph{event conditions} are specified by a function $ec$, assigning
to each event a \emph{guard} or \emph{activation condition} (a
boolean predicate depending on system variable, stating when a
transition can fire) and a \emph{reset} (specifying how variables
are modified by the event). For example, $\ec(\ev{full}) = (B =
max_B,true)$ states that the uplink is shut down when the buffer
reaches its maximum capacity $max_B$, and no variable is modified.
If we wanted to model a policy throwing away a fraction $\rho$ of
the packets when the buffer becomes full, then we could have defined
$\ec(\ev{full}) = (B = max_B,B' = (1-\rho)B)$. Deterministic events
in HYPE are \emph{urgent}, meaning that they fire as soon as their
guard becomes true.

Stochastic events $\sev{a}\in \Ev_s$ have an event condition composed
of a stochastic rate (replacing the guard of deterministic events)
and a reset. For instance, $\ec(\sev{on_{in}})  =  (k_{in}^{on},true)$
states that the reception of packets is a stochastic event happening
at times exponentially distributed with constant rate $k_{in}^{on}$.
In general, rates define exponential distributions and can be
functions of the variables of the system.

For completeness, we provide the  formal definition of the syntax
of stochastic HYPE.

\begin{definition}
A \emph{stochastic HYPE model} is a tuple 
$(\ConSys,\V,\IN,\IT,\Ev_d,\Ev_s,\Ac,\ec,\iv,\EC,\ID)$ where
\begin{itemize}
\item $\ConSys$ is a controlled system as defined below.
\item $\V$ is a finite set of variables.
\item $\IN$ is a set of influence names and $\IT$ is a set of influence
type names.
\item $\Ev_d$ is the set of instantaneous events of the form $\ev{a}$ and $\ev{a}_i$.
\item $\Ev_s$ is the set of stochastic events of the form  $\sev{a}$ and
$\sev{a}_i$.
\item $\Ac$ is a set of activities of the form
$\alpha(\W) = (\iota,r,I(\W)) \in (\IN \times
\mathbb{R} \times \IT)$ where $\W \subseteq \V$.
\item $\ec:\Ev \rightarrow \EC$ maps events to event conditions. Event
conditions are pairs of activation conditions and resets. Resets
are formulae with free variables in $\V \cup \V'$. Activation
conditions for instantaneous events $\Ev_d$ are formulas  with
free variables in $\V$, while for stochastic
events of $\Ev_s$, they are functions
$f:\bbR^{|\V|}\rightarrow\bbR^+$.
\item $\iv: \IN \rightarrow \V$ maps influence names to
variable names.
\item $\EC$ is a set of event conditions.
\item $\ID$ is a collection of definitions consisting
of a real-valued function for each influence type name
$\assign{I(\W)} = f(\W)$ where the variables in
$\W$ are from $\V$.
\item $\Ev$, $\Ac$, $\IN$ and $\IT$ are pairwise disjoint.
\end{itemize}
\end{definition}

\begin{definition}
A \emph{controlled system} is constructed as follows.
\begin{itemize}
  \item \emph{Subcomponents} are defined by $C_s(\W) = S$, where $C_s$
  is the \emph{subcomponent name} and $S$ satisfies the grammar
  $S'::= \ssev{a}:\alpha.C_s\ |\ S'+S'$
  ($\ssev{a}\in\Ev=\Ev_d\cup\Ev_s$, $\alpha\in\Ac$), with the
  free variables of $S$ in $\W$.
  \item \emph{Components} are defined by $C(\W) = P$, where $C$
  is the \emph{component name} and $P$ satisfies the grammar
  $P'::= C_s(\W)\ |\ C(\W)\ |\ P'\sync{L}P'$, with
  the free variables of $P$ in $\W$ and $L\subseteq\Ev$.
  \item An \emph{uncontrolled system} $\Sigma$ is defined according
  to the grammar $\Sigma' ::= C_s(\W)\ |\ C(\W)\ |\
  \Sigma'\sync{L}\Sigma'$, where $L\subseteq\Ev$ and $\W \subseteq \V$.
\item \emph{Controllers} only have events:
$M ::= \ev{a}.M\ |\ 0 \ |\ M + M$ with $\ev{a} \in \Ev$ and $L
\subseteq \Ev$ and $\Con ::= M\ |\ \Con \smash{\sync{L}} \Con$.
\item A \emph{controlled system} is
$\ConSys ::= \Sigma \smash{\sync{L}} \ev{init}.\Con$ where $L
\subseteq \Ev$. The set of controlled systems is $\CSys$.
\end{itemize}
\end{definition}

The semantics of HYPE has been defined in \cite{GalpGBH:09a,HYPE-journal},
where a mapping to Hybrid Automata \cite{HenzHH:95a} is also
discussed.  The semantics of stochastic HYPE \cite{BortBGH:11a},
instead,  is given  in terms of TDSHA (Transition-Driven Stochastic
Hybrid Automata, \cite{BortBP:09a}), which are a high level
representation of PDMPs (Piecewise Deterministic Markov Processes,
\cite{Davis93}).  PDMPs are stochastic hybrid processes which
interleave a deterministic evolution, described by a set of
differential equations (depending on the current discrete mode of
the system), with discrete jumps, which can be of two types:
\emph{spontaneous}, happening at exponentially distributed random
times, and \emph{forced}, happening when specific conditions on
system variables are met. Both kind of events can reset the state
of the system according to a specified reset policy. Intuitively,
the dynamics of a stochastic HYPE model is as follows: the system
variables will evolve following the solution of a set of ODE, defined
by the influences active in the system, one for each influence name.
The events that can happen, instead, are determined by the current
state of the controller.  Active stochastic events happen at random
times, while deterministic events happen when their guard becomes
true. In both cases, the reset policy of the event is applied.
Moreover, the state of the controller and the set of active influences
is updated according to model structure. In particular, all influences
preceded in subcomponents by the  event that occurred will replace
the ones with the corresponding name, so that the continuous dynamics
can have different modes of operation.

All HYPE models that will be considered in the paper comply with
the definition of well-defined HYPE models, given in~\cite{HYPE-journal}.
Essentially, each subcomponent must be a self-looping agent of the
form $S = \sum_{i=1}^k \ssev{a}_i\pc\alpha_i.S + \ev{init}\pc\alpha.S$,
with each $\alpha_i$ of the form $(\iname{i}_S,r_i, I_i)$, where
$\iname{i}_S$ is an influence name appearing only in subcomponent
$S$. Furthermore, synchronization must involve all shared events.

\subsection{Simulation software}
\label{sec:tool}

In this section, we provide some details about the implementation
of stochastic HYPE that we used to analyse the model of an opportunistic
network presented in the paper. This software tool supports an
automatic extraction of basic statistics from a set of stochastic
runs, and has plotting facilities (including 3D plots and distribution
histograms) and data export facilities. Furthermore, it supports
automatic exploration of the parameter space. The user interface
is command-line-based, and uses a simple script language to instruct
the software.

\begin{figure}[t]
\begin{center}
\hspace*{1.5cm}
\begin{minipage}{15cm}

{\scriptsize
\begin{verbatim}
hype model network_node

#definitions
var B = 0;              //buffer size
param maxB = 100;       //buffer capacity
param r_in = 1;         //input rate
param r_out = -2;       //output rate
param kon_in = 0.5;     //uplink activation rate
param koff_in = 0.05;   //uplink deactivation rate
param kon_out = 0.02;   //downlink activation rate
param koff_out = 0.01;  //downlink deactivation rate

function const() = 1;        //constant function	
guard above(X,K) = X >= K;   //X >= K
guard below(X,K) = X <= K;   //X <= K

#mappings
infl in :-> B;       //input influence
infl out :-> B;      //output influence
event on_in   =  :-> @ kon_in;         //input activation
event off_in  =  :-> @ koff_in;        //input deactivation
event on_out  =  :-> @ kon_out;        //ouput activation
event off_out =  :-> @ koff_out;       //output deactivation
event full    =  above(B,maxB) :-> ;   //buffer full
event empty   =  below(B,0)    :-> ;   //buffer empty

#subcomponents
//template to define a switch between two states
switch(on,off,block,r) := off,block,init:[0,const()] + on:[r,const()];

#components
input  := switch(on_in,off_in,full,r_in):in;       //input component
output := switch(on_out,off_out,empty,r_out):out;  //output component
sys := input <*> output;                           //uncontrolled system

#controller
con_in := on_in.con_in_1; con_in_1 := off_in.con_in + full.con_in;         //input controller
con_out := on_out.con_out_1; con_out_1 := off_out.con_out + empty.con_out; //output controller
con := con_in || con_out;                                                  //system controller

#system
sys <*> con;     //system
\end{verbatim}
}

\end{minipage}
\end{center}
\caption{Code for the example given in Figure \ref{simpleNetwork}.}
\label{fig:simhya}
\end{figure}
\label{sec:Opportunistic}

The basic idea behind the implementation, written in Java and available
on request from the authors, is to flatten a HYPE model into a
representation in which additional \emph{discrete} variables (i.e.,
variables that can take only a finite set of values) are introduced
to keep track of the current active influences and the current
states of the controller. The number of variables required for this
encoding is easily seen to be linear in the size of the system, as
it requires an additional number of variables equal to the number
of different subcomponents plus the number of different states of
the controller. Furthermore, this approach has the advantage of
avoiding an explicit construction of all the modes of the (stochastic)
hybrid automaton associated with a HYPE model. This is possible
since modes of such an automaton are uniquely identified by the
values of the discrete variables introduced.

To illustrate how the encoding works in terms of the generated ODEs,
consider the $\Input$ component of Section \ref{sec:sHYPE}.  To
model which influence is active between $\alpha_1 = (in,r_{in},\const)$
and $\alpha_0 = (in,0,\const)$, we need a new variable, $I_{\Input}$,
taking values in $\{0,1\}$. If $I_{\Input}$ equals zero, the active
influence is $\alpha_0$, otherwise the active influence is $\alpha_1$.
This means that the component of the ODEs associated with variable
$B$ generated by $\Input$ is of the form $r_{in}\assign{const}\langle
I_{\Input} = 1\rangle + 0\assign{const}\langle I_{\Input} = 0\rangle
=  r_{in}\langle I_{\Input} = 1\rangle$, where $\langle \cdot
\rangle$ denotes the logical value of a boolean predicate expressed
as $0$ or $1$. Resets and guards of events are modified in order
to correctly update the discrete variables introduced. For instance,
the reset of event $\ev{full}$ resets $I_{\Input}$ to 0.

The tool provides simulation of stochastic and non-stochastic HYPE
models, and uses the Java mathematical library MathCommons
\cite{mathcommons} to numerically integrate the ODEs, exploiting
its embedded event detection system to manage the firing of events.
In particular, stochastic simulation is dealt with in the following
way \cite{Wilkinson,BortBP:09a}. Consider a stochastic event with
rate $\lambda(t)$, depending on time via the continuously evolving
variables of the system. We compute its cumulative rate $\Lambda(t_0,t)
= \int_{t_0}^t \lambda(s)ds$ by coupling with the ODE system, the
following equation for $\Lambda$: $\smash{\frac{d\Lambda(t_0,t)}{dt}
= \lambda(t)}$, with $\Lambda(t_0,t_0) = 0$. Then, we fire the
stochastic transition as soon as $\Lambda(t_0,t) = -\ln(U)$, where
$U \sim \mathit{Unif}(0,1)$ is a uniformly distributed random number
in $[0,1]$, sampled using the pseudo-number generator of MathCommons
library. Notice that if the rate $\lambda(t) = \lambda$ is constant,
then the firing time is $-\frac{1}{\lambda}\ln(U)$, and hence we
have the standard Monte Carlo inversion method to simulate exponentially
distributed random variables \cite{Wilkinson}.

We conclude this section with some details of the language supported
by the tool to model with HYPE.  Each HYPE model consists of 6
sections. The \verb"#definitions" section contains the definition
of system variables, parameters, expressions shorthands, user-defined
functions (which replace influence types in the tool) and boolean
predicates.  The \verb"#mappings" section is devoted to the  definition
of influences (mapping them to variables) and  events (specifying
their name, guard, reset, and, for stochastic events, rate). In the
tool, stochastic events can be guarded. This is rendered in HYPE
using suitable discontinuous rate functions.  The \verb"#subcomponent"
section contains the description of subcomponents, which can be
parameterised (with respect to variables and events) in order to
reuse the same definition more than once.  In particular, the user
can assign more than one event to the same influence and the influence
name is assigned to the whole subcomponent, in order to comply to
the restrictions of well-defined HYPE models.  The \verb"#component"
section, instead, contains the (parametric) description of system
components, including the uncontrolled system.  The \verb"#controller"
section contains the definition of sequential and compound controllers.
Finally, the \verb"#system" section specifies the complete system
by combining a controller and an uncontrolled system. The code for
the example of Figure \ref{simpleNetwork} is given in
Figure~\ref{fig:simhya}.  We would like to add additional
parameterisation abilities to the software to support systems where
we define many similar components. We address this point further
in Section~\ref{sec:furtherResearch}.

\section{Opportunistic networks}

Since stochastic HYPE allows for the modelling of discrete quantities in
a fluid manner, it is suitable for network modelling. Furthermore, it
has stochastic aspects that model randomness, making it appropriate to
model the disconnectedness that can happen in opportunistic networks.
Networks are opportunistic when nodes can communicate
even though there may never be a direct path between them. They use a
\emph{store-carry-forward} approach and decisions
about routing are determined dynamically with policies based on the
notion of getting a packet closer to its final destination
\cite{PeluPPC:06a}. Delays may occur but networks of this type can
be deployed in environments where disconnectedness is possible but
increased time for packet delivery is acceptable.
The major challenges in such networks
\cite{PeluPPC:06a,HuanHLT:08a,ShenSMC:08a}
include the following.
\begin{description}
\item[Disconnectedness:] A direct path may occur very
infrequently or never between any two nodes in the network.
\item[High latency and low data rate:] Due to disconnectedness,
there can be significant delays for an individual packet, including
those caused by queueing at an intermediate node. This obviously
can result in low throughput.
\item[Limited resources:] Nodes are often battery-driven and hence need
to conserve energy to lengthen their lifetimes. This means that the
amount of storage space or strength and length of radio usage for
communication are limited.
Additionally, nodes may become permanently disabled
due to a hostile environment.
\end{description}
Various protocols have been designed to mitigate the problems caused
by these challenges and these protocols have specific objectives
\cite{ShenSMC:08a}. The main objective is to maximise the probability
of a packet reaching its destination. Ideally, at the same time,
both the delivery delay and the resource usage should be minimised.
Storage capacity at nodes should be sufficient, both to cope with
the inherent latency and in certain cases, to store copies of
messages that could be lost.

There are different ways to categorise routing/forwarding protocols:
deterministic versus stochastic \cite{ZhanZ:06a}, with or without
infrastructure \cite{PeluPPC:06a} or most commonly, flooding versus
forwarding \cite{HuanHLT:08a,ShenSMC:08a}. In flooding protocols,
packets are forwarded to many nodes. Variations include epidemic
routing which is based on a model of disease transmission, where
nodes that have ``recovered" do not forward packets they have already
seen \cite{VahdVB:00a}. Additional conditions can be added to
epidemic routing to reduce resource usage. Other flooding approaches
involve the estimation of the probability of delivery by a node,
and this is used in deciding which nodes packets should be forwarded
to. An example is PROPHET \cite{LindLDS:03a}.

In forwarding protocols, a single copy of a packet moves through
the network. Decisions about which node is the best node to move
to can be done by location (how close the node is to the destination
node), knowledge about the network provided by oracles \cite{JainJFP:04a}
or other characteristics of the network that can be obtained by a
node through interaction with other nodes. For example, the MaxProp
protocol is based on historical data of path likelihood
\cite{BurgBGJL:06a}.

In the context of our case study, two wildlife monitoring projects
are of interest: ZebraNet \cite{JuanJOWMPR:02a} and SWIM
\cite{SmalSH:03a}. In the first case, zebra are fitted with collars
and the data is collected by a mobile node on a vehicle that moves
around the area. In second case, whales are tagged with sensors.
In both cases, flooding can be used as the protocol. In flooding,
whenever two animals are in sufficiently close proximity, they
exchange data.  In this way, assuming one animal comes close enough
to the mobile nodes or base stations, there is sufficient proximity
between animals, and no data buffers become full, all data will
arrive at the base station. In the case of ZebraNet, the history-based
protocol is also used. Here, each sensor keeps a value which gives
an indication of when it last interacted with the mobile node. When
deciding which neighbour sensor to send data to, the one with the
highest value is chosen.  This protocol has been shown to outperform
the flooding protocol.

ZebraNet has similarities with carrier-based routing (which is
classified by \cite{PeluPPC:06a} as routing with mobile infrastructure).
In these protocols \cite{JainJSBBR:06a,ZhaoZAZ:04a}, particular
mobile nodes which can be called carriers, supports, forwarders,
mules or ferries, collect data from other, possibly stationary,
nodes. In some protocols, only the ferries collect data and in
others, non-ferry nodes exchange data as well. Our case study is
based on the former.

\subsection{Our case study}

As an initial test for our modelling of opportunistic networks, we
have chosen a relatively simple scenario. Since our fluid packet
approach is most useful in cases where there are large amounts of
data, we consider an example where video data is captured by
stationary sensors. The idea is that they are motion activated with
a low number of activations expected each day. Because these sensors
are required to run off battery, it is not desirable for them to
have powerful enough radio to share data over distance. Hence, a
vehicular ferry moves around the area in which the video sensors
are located and returns to a base station where the ferry delivers
the data. This scenario could occur in a wildlife reserve or any
scenario where video data is to be collected, but is not required
in real-time. We do not assume that the primary purpose of the
vehicle involved is to collect data from the nodes. It could be
involved in supply delivery or game viewing, but we do assume some
flexibility in routing as we consider later.

If we assume a fixed disk size for the video sensors, and no restriction
on the amount of data that can be delivered to the base station, then
the parameters of interest relate to the ferry and include buffer size,
route taken and speed of movement. In the next section, we discuss how
this can be modelled.

\section{A framework for modelling opportunistic networks}
\label{sec:Model}

The basic element of our model is the network node. To be able to model
opportunistic networks in the most general way, we assume that each node has a
fixed buffer capacity, and that it can keep track of multiple streams of data.
These streams could represent data with different priorities, data with
different destinations or different types of data; or combinations of
priority, destination and type.

A node should be able to accept input; offer output; discard data
from the buffer, either to free up space by dropping current data
or to remove stale data, and generate data. Moreover, it should be
able to keep track of the total data it has dropped, input or
generated. It should also be able to make decisions about what data
to input (how much, which and from whom), output (how much, which
and to whom) and drop (how much and which). In certain instances,
it may also need to make decisions about what data to generate.

To model this node in stochastic HYPE, we require variables to
capture the current buffer level for each stream, together with the
lifetime input and generated data for each stream, and lifetime
dropped data for each stream.  This gives us variables $\Level_{i,v}$,
$\TotalIG_{i,v}$ and $\TotalD_{i,v}$ for node $i$ and stream $v$.
Clearly, the value of the variables for everything except buffer
levels will not decrease.

Each node has subcomponents for input, output, generation, drop and
removal (where the second last term refers to discarding current
data and the last term to discarding stale data). It also has two
further subcomponents to keep track of data input and data generated.
We have the following HYPE components for each node $i$ and each
stream $v$. Here, the symbol $\syncstar$ indicates synchronisation
on all shared events.

\medskip
$\begin{array}{rcl}
\hspace*{-0.2cm}
\Node_{i,v} & \rmdef & \Input_{i,v} \syncstar \Output_{i,v} \syncstar
\Gen_{i,v} \syncstar \Rem_{i,v} \syncstar \Drop_{i,v} \syncstar
\KeepI_{i,v} \syncstar \KeepG_{i,v} \\
\\
\end{array}$

The influences that appear in $\Input_{i,v}$, $\Output_{i,v}$,
$\Gen_{i,v}$ and $\Rem_{i,v}$ are mapped to the variable $\Level_{i,j}$, those in
$\KeepI_{i,v}$ and $\KeepG_{i,v}$ are mapped to the variable $\TotalIG_{i,j}$ and that
in $\Drop_{i,v}$ to the variable $\TotalD_{i,j}$.

Each node has a controller for the first five subcomponents.
Controllers are not required for the last two subcomponents since
the events for these subcomponents appear in the controllers for
input and generation.  Controllers are required for each stream as
they may need to be treated separately, for example in the case of
one stream being prioritised over another.

\medskip
$\begin{array}{rcl}
\CNode_{i,v} & \rmdef & \CInput_{i,v} \syncstar \COutput_{i,v} \syncstar
\CGen_{i,v} \syncstar \CRem_{i,v} \syncstar \CDrop_{i,v} 
\end{array}$
\medskip

The example in Section~\ref{sec:sHYPE} provides a very simple version
of such a node which only has input and output capability. The
controllers must deal with aspects such as full and empty buffers,
as well as switching between different functions, for example
switching between input on and input off. 

The next important aspect to model is the interaction between nodes
in the network. Each possible connection between two nodes (some
nodes may never have the ability to connect) has a controller that
synchronises on proximity and brings up the link, takes down the
link at the end of proximity or due to any other condition that
could cause the link to end, and then does some housekeeping. Hence,
we define $\CLink_{i,j}$ the controller for the link between nodes
$i$ and $j$.

Then for each stream of data, there is a controller that involves
events that determine whether data exchange should happen for that
stream and whether the link should be uni- or bidirectional. The
directionality of the link depends both on the characteristics of
the modelled system and the policies used. This controller also includes
a housekeeping subcontroller that synchronises with the housekeeping of
the controller of the link between the two nodes.

\medskip
$\begin{array}{rcl}
\CStream_{i,j,v} & \rmdef & \CUni_{i,j,v} \syncstar \CUni_{j,i,v} \syncstar
CBi_{i,j,v} \syncstar \CTidy_{i,j,v} 
\end{array}$
\medskip

Finally, there are controllers which model the proximity of nodes.
This is currently done abstractly, using a random variable to capture
delays between connectedness.  A recent paper describes the expected
meeting time between nodes for various mobility models such as
random direction, random waypoint and community-based (both for
homogeneous and heterogeneous nodes) and suggest that these expected
values can be used as rates to describe exponential distributions
\cite{SpyrSPR:06a}. This allows us to model these specific
types of mobility in this abstract fashion. However, as future work,
we plan to develop more concrete models that describe movement in
two-dimensional space, since HYPE can model this type of continuous
behaviour in a straightforward way.

Hence, to construct an opportunistic network model in stochastic
HYPE, we need to define a number of nodes based on our template,
the appropriate connection controllers and proximity controllers,
and most importantly, the policies and protocols that will
be used in the network. Our longer-term goal is to develop a front
end that will allow a user to specify nodes, connections, policies
and protocols from which a HYPE model will be generated. This will
provide a simulation tool for networks that should be faster than simulators
that trace every packet, when the number of packets is large.

\subsection{Our case study}

For our specific case study, we used the basic node we have developed.
Video sensor nodes require generation, discard and output capabilities;
the ferry requires input and output, and the base station, input
only.  Possible links between nodes cover upload from sensors to
the ferry and upload from the ferry to the base station. There are
also two different proximity controllers: one allows the ferry to
have a random route between nodes, and the other imposes a fixed
route that is cyclically repeated.

In our case study, the policies are straightforward -- unidirectional
communication between a sensor and the ferry is set up 
only if the ferry is not currently communicating with another
sensor. However, in a scenario with more than one ferry, it would
be possible to implement a policy allowing a sensor node to choose
which ferry to upload data to, making a decision based on
certain ferry characteristics. Since we are dealing with a system
where large amounts of data are generated, it makes sense to use
this protocol rather than any other. In this scenario, it would be
problematic to use flooding as the system is data bound, and hence
an excessive transmission and storage of data cannot be recommended.
\subsection{Results}
\label{sec:Results}

\begin{figure}[t]
\begin{center}
\includegraphics[height = 8.5cm]{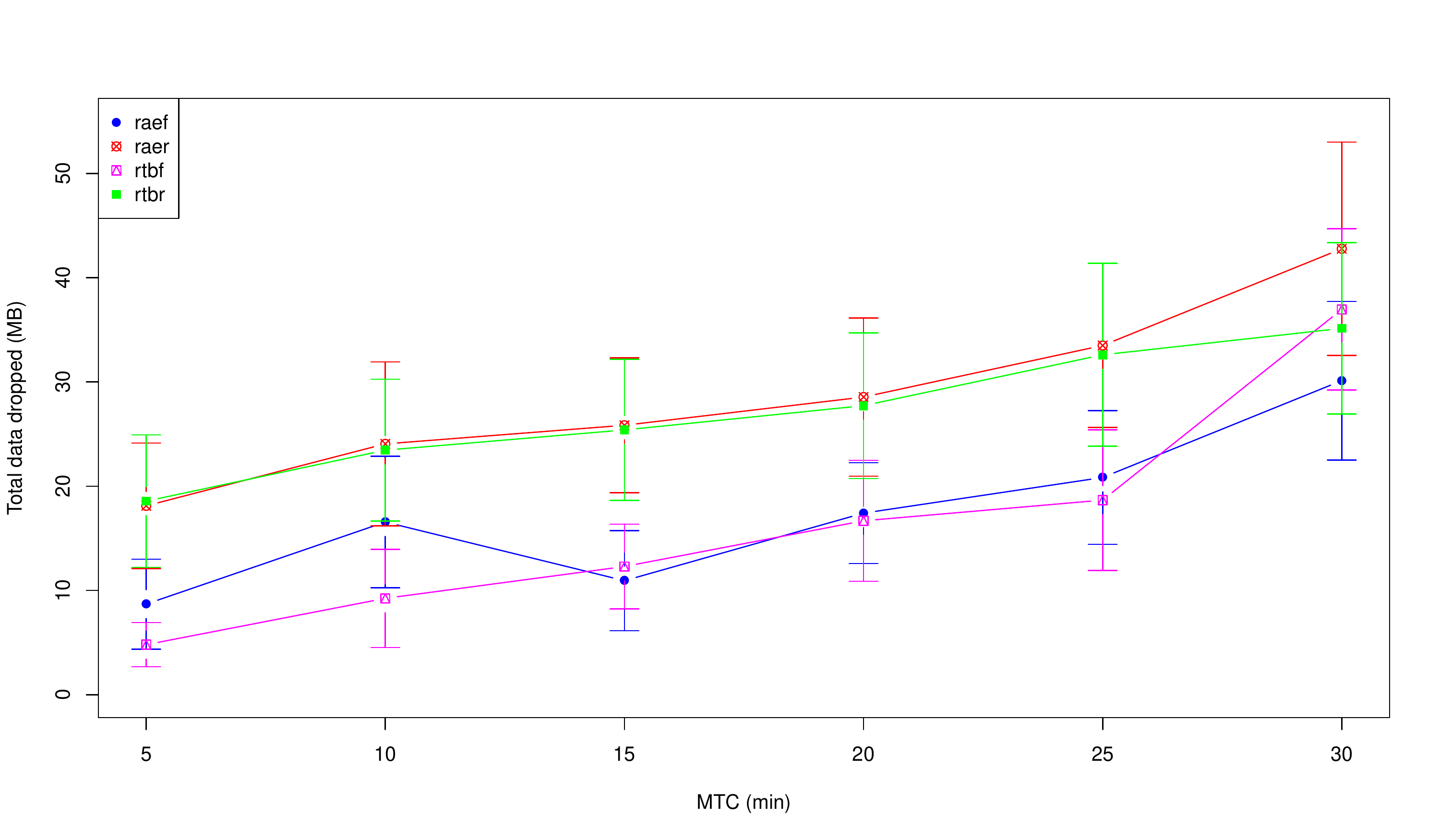}
\end{center}
\vspace*{-0.7cm}
\caption{Average (200 simulation runs) data dropped for different mean
time to contact ($\MTC$) values. The bars delimit the 95\% confidence
interval. See the text for the description of the
four different scenarios.}
\label{fig:graphs-MTCdrop}
\end{figure}

The specifics of our case study are as follows: we assume that there
are 10 video sensors and that the ferry only collects data during
an 8 hour period, and we are interested both in what buffer size
the ferry requires and how often there is contact between the ferry
and a sensor (described by the mean-time-to-contact variable $\MTC$),
which is effectively the speed of the ferry.

Each video sensor has 250MB of disk space, and on average will
record video three times a day for an average of 3 minutes each
time; the video will require 10Mb for each minute.  The upload speed
from sensor to ferry is 1MB/s and the upload speed from ferry to
base station is 30MB/s.

We consider 4 different scenarios. For \emph{raer}, the ferry only
returns to base at the end of the 8 hour period and has a random route.
In the case of \emph{raef}, the return to base is also at the end but
there is a fixed route. For \emph{rtbr} and \emph{rtbf}, the ferry
returns to base whenever it is full (incurring a penalty of extra
distance to travel) and there are random and fixed routes, respectively.
In the experiments to explore different values for mean time to
contact, the ferry buffer size was set to 1000MB. In the experiments
to consider different ferry buffer size the mean time to contact
was set to 15 minutes. Each simulation took around 6.5 seconds on a
standard laptop.

\begin{figure}[t]
\begin{center}
\includegraphics[height = 8.5cm]{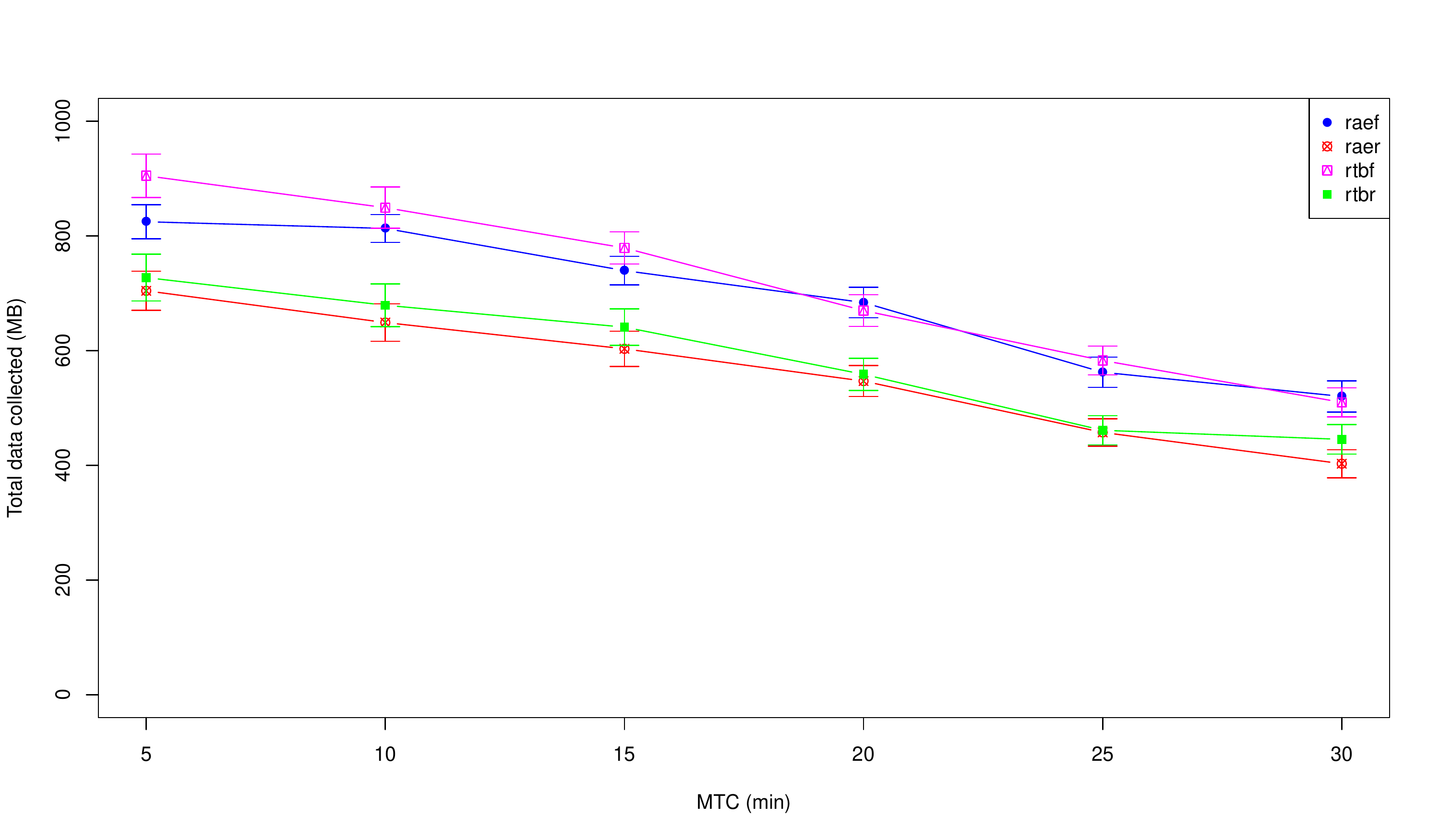}
\end{center}
\vspace*{-0.7cm}
\caption{Average (200 simulation runs) data collected by ferry for
different mean time to contact ($\MTC$) values. The bars delimit the
95\% confidence interval.}
\label{fig:graphs-MTCcollect}
\end{figure}

When we consider data dropped versus mean time to contact (see
Figure~\ref{fig:graphs-MTCdrop}), as mean time to contact increases
(which means visits are less frequent) the amount of data dropped
increases, which is what we would expect. In general, it can be
seen that the fixed routes (\emph{raef} amd \emph{rtbf}) result in
less data drop, which can be ascribed to fairness, in that each
node get its turn whereas in the random case, it may not get a turn
at all.
Similarly, Figure ~\ref{fig:graphs-MTCcollect} show that as frequency
of visits decreases, the amount of data collected by the ferry
decreases, and the fixed routes (\emph{raef} amd \emph{rtbf}) perform
better in collecting more data.

\begin{figure}[t]
\begin{center}
\includegraphics[height = 8.5cm]{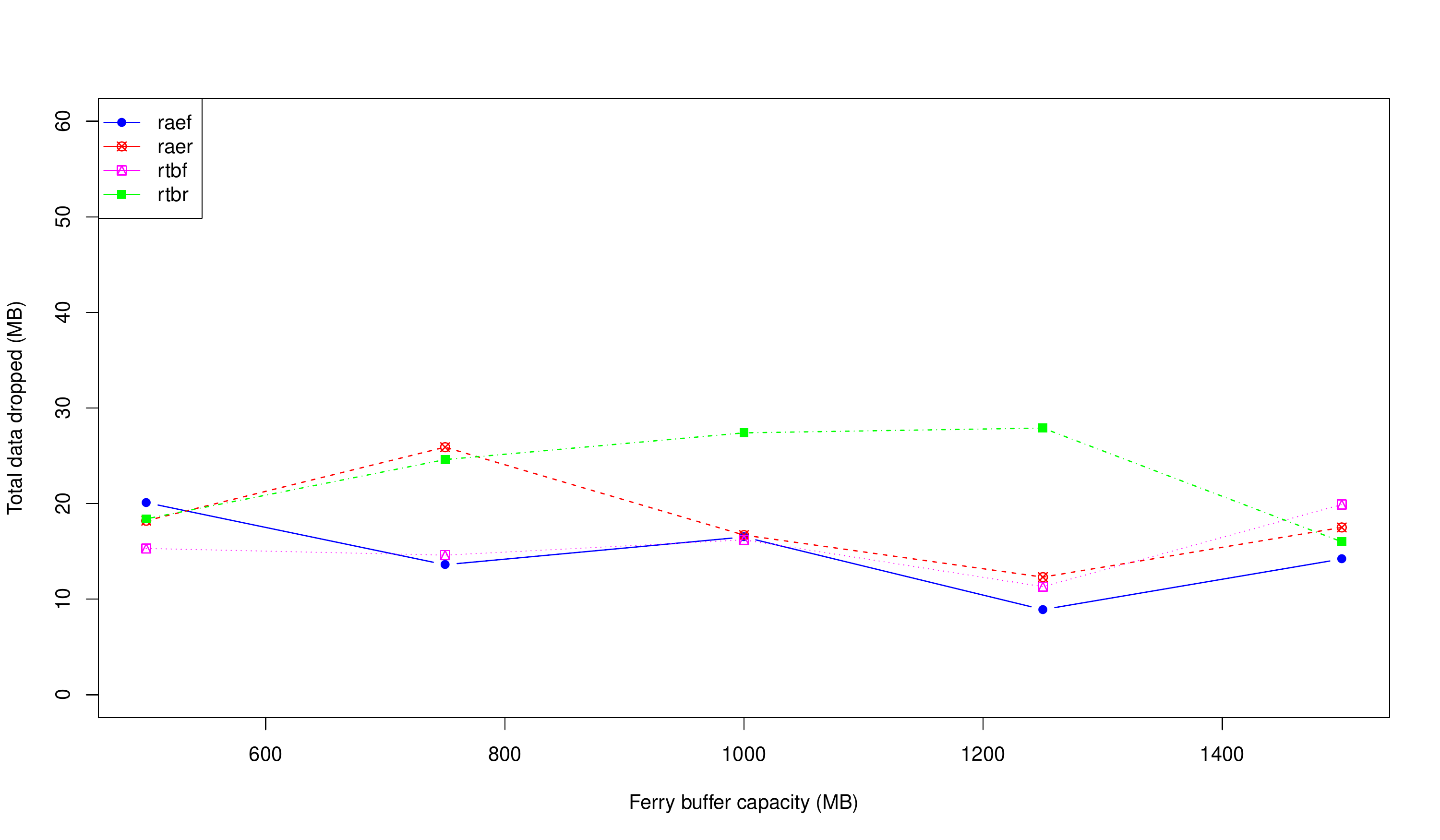}
\end{center}
\vspace*{-0.7cm}
\caption{Average (100 simulation runs) data dropped for different ferry
buffer sizes}
\label{fig:graphs-mcpdrop}
\end{figure}

\begin{figure}
\begin{center}
\includegraphics[height = 8.5cm]{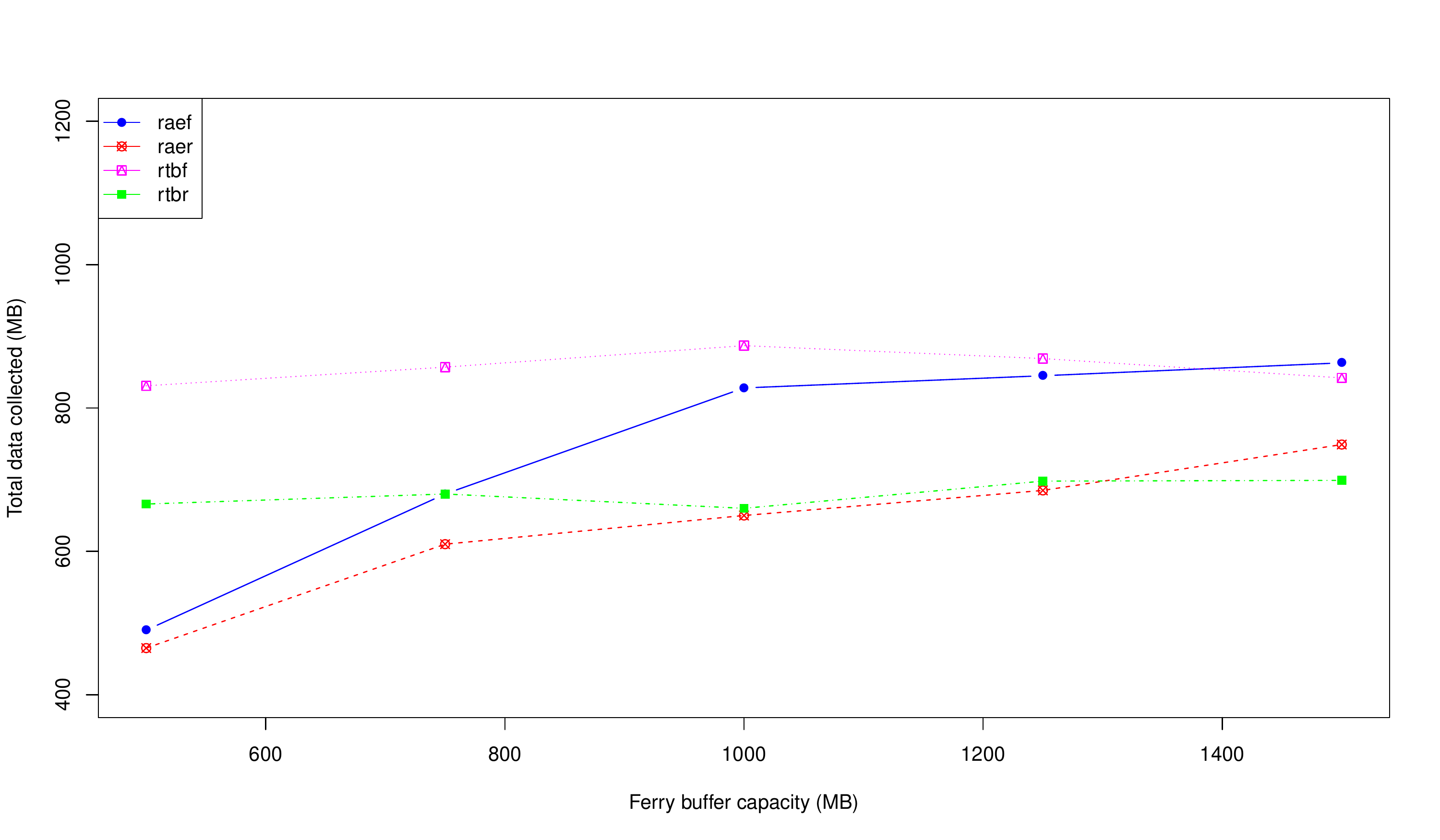}
\end{center}
\vspace*{-0.7cm}
\caption{Average (100 simulation runs) data collected by ferry for
different ferry buffer sizes}
\label{fig:graphs-mcpcollect}
\end{figure}

For the ferry buffer size, we found that there appeared to be limited
correlation between the different protocols and amount of data
dropped, shown in Figure~\ref{fig:graphs-mcpdrop}. By contrast,
in Figure~\ref{fig:graphs-mcpcollect}, the amount of data collected
by the ferry, is understandably reduced for low buffer capacities
in the case where the ferry does not return to base when full
(\emph{raer} and \emph{raef}). At the higher buffer capacities, again
the fixed routes out perform the random routes.

This case study illustrates how stochastic HYPE can be used
to model these networks. It is relatively straightforward to add
further nodes of all types, and hence model a larger system of the
same type.

\section{Related work}
\label{sec:relatedWork}

Other hybrid process algebras to describe hybrid systems include
$\smash{\mathit{ACP}_{hs}^{srt}}$ \cite{BergBM:05a}, hybrid
$\smash{\Chi}$ \cite{BeekBMRRS:06a}, $\phi$-calculus \cite{RounRS:03a}
and HyPA \cite{CuijCR:05a}.  The aspect of HYPE that distinguishes
it from these process algebra, is that in HYPE the ODEs emerge from
the semantics and are not required to be specified monolithically
in the syntax because of the use of individual flows in HYPE.
Additionally, unlike the process algebras mentioned and hybrid
automata, it is possible to combine two HYPE models where a variable
can be shared between models since it is possible to combine all
the influences that apply to this shared variable. A more detailed
comparison between HYPE and other hybrid modelling formalisms can
be found in~\cite{GalpGBH:09a,HYPE-journal}. To the best of our
knowledge, no hybrid process algebra has previously been used to
model networks using a fluid packet approach.  Recently, an approach
based on rewriting logic, Hybrid Interacting Maude, has been developed
\cite{FadlFOA:11a} but to date this research has focussed on thermal
systems.

Other formal approaches to performance modelling of opportunistic
networks have appeared in the literature in recent years.  Much
work has focussed on modelling the mobility patterns of nodes within
the network, a feature that clearly has a strong impact on the
performance that can be achieved.  Examples include \cite{Xu:2009}
and \cite{SpyrSPR:06a} in which the authors analyse the expected
meeting time for various mobility models and bounds on delays.
Other papers focus on the performance measures such as message delay
and compare, as we do, the routing policies which may be applied.
For example, Picu and Spyropoulos \cite{Picu:2010} use expensive
Markov Chain Monte Carlo simulation to assess optimal relay selection
for multicast communication in opportunistic networks, while
\cite{KeunKLZ:11a} presents analytic bounds on message delivery
capacity.  Another example \cite{Pass11} considers the provisioning
of a network in order to minimize the delay using an analytical
model based on queueing theory.  Their framework is more general
than ours in the sense that services, rather than simply messages,
are exchanged opportunistically between nodes and results as well
as service requests are also exchanged.   However, it should be
possible to extend our modelling framework to encompass this richer
scenario.  Closest to our work in terms of formality is the work
of Garetto and Gribaudo, but this presents a purely discrete model
in terms of a state-labelled Markov chain which is subjected to
probabilistic model checking \cite{GareGG:06a} and is therefore
limited in the size of system which can be considered.

Other simulators for opportunistic networks have been proposed, for
example the ONE \cite{KeraKKO:10a} and a virtual test platform
\cite{DeepDTN:08a}. These simulators work at the packet and message
level and do not introduce a fluidisation of data flow. Additionally,
examples studied quite often consider generation rates as low as a
message an hour or a day. By using a fluid approach, we can model 
much higher generation rates.

Lastly, we mention other fluid approaches to modelling networks
(necessarily incomplete due to space constraints).  These
include simulation \cite{KiddKSWU:03a}, fluid stochastic Petri nets (FSPNs)
\cite{HortHKNT:98a,TuffTCT:01a,AjmoAMGMS:01a,GaetGGMS:05a}
and mean field approximations \cite{BenaBLB:08a,BakhBCFH:11a}. These approaches,
as far as we know, have not been applied to opportunistic networks and
they do not offer the compositionality that a process
algebra provides. 

Petri net approaches include modelling a single cell of a wireless
internet access system as a deterministic and stochastic Petri net;
a generalised stochastic Petri net; and a FSPN \cite{AjmoAMGMS:01a}.
The FSPN model is much faster to solve than the other two, and has
good accuracy apart from a few specific scenarios. The FSPN model
has a single fluid place to represent the buffer of the system.
Another Petri net approach presents a FSPN model of client-server
interaction in a peer-to-peer network \cite{GaetGGMS:05a} which is
used to obtain distributions describing file transfer times. The
single fluid place represents the download of a file. In both these
examples, there is no flow between continuous places, and it is not
clear how these models could be extended to model systems of multiple
cells, or multiple clients and servers where continuous flows occur
betweeen different elements of the system, in contrast to our approach.

\section{Further research and conclusions}
\label{sec:furtherResearch}

In terms of future work, we plan to improve the software tool in
at least two directions. First, we want to improve the modularisation
of the input language, allowing the modeler to write parametric
templates corresponding to generic system components like ferries,
data stations, and connections. Secondly, we plan to implement in
the hybrid simulation more clever management of the discrete variables
denoting modes of the automaton, using data structures such as
dependency graphs to reduce the amount of times each guard of a
discrete transition is tested.  Preliminary work on this (together
with bytecode on-the-fly compilation of mathematical expressions)
has shown a thirty-fold increase in performance over the execution
time given in Section~\ref{sec:Results}.  In addition, we plan to
implement a multithread support to exploit multi-core processors
to reduce simulation time.

As for the opportunistic networks, exploiting modularisation of
HYPE code, we plan to define a library of components and an higher
level graphical interface to construct a model. As mentioned in
Section~\ref{sec:Model}, we wish to investigate the use of
two-dimensional models of node movement.  Furthermore, we plan to
explore the use of HYPE bisimulations to reduce model size when
possible. In certain cases, it may also be possible to reduce a
sequence of events to one event while retaining the same behaviour
with respect to simulation. As mentioned in Section~\ref{sec:Results},
it should be easy to increase the size of the networks to be modelled,
and hence we will investigate how our approach scales and what
limitations there might be.

As a general principle, hybrid simulation can be more efficient
than discrete event simulation (without continuous flows) as long
as the computation time for solving the ODEs is shorter than the
computation time for simulating the events that the fluid approach
removes. We wish to establish whether this principle holds in
modelling opportunistic networks and what the trade-off is between
efficiency and accuracy of the modelling. Currently, we have focussed
on providing averages and standard deviations over a number of
simulation runs. An alternative approach is to formally specify the
properties we are interested in using a stochastic logic and then
to apply statistical model checking.

To conclude, we have presented a general framework for constructing
models of opportunistic networks using stochastic HYPE. We have
illustrated this through a case study of a ferry that collects data
from video nodes and delivers it to a base station. The results of
our simulations of the model show, as we would expect, that fixed
routes are likely to be more fair and have less data dropped and
more data collected. Additionally, at low buffer sizes, the penalty
of returning to the base is justified to avoid dropping data
unnecessarily.

\vspace*{-0.3cm}
\paragraph{Acknowledgements:\ }
This research is supported by Royal Society International Joint
Project JP090562.

\bibliographystyle{eptcs}
\bibliography{qapl2012}

\end{document}